\documentclass[twocolumn,preprintnumbers,amsmath,amssymb,super]{revtex4}
\usepackage{graphicx}
\usepackage{bm}
\DeclareGraphicsExtensions{.pdf,.png,.jpg}
\usepackage{epstopdf}

\begin{document}


\title{Stoner factors of doped 122 Fe-based superconductors: First principles results}
\author{Smritijit Sen$^{1,2}$, Haranath Ghosh$^{1,2}$}
\affiliation{$^1$Homi Bhabha National Institute, Anushaktinagar, Mumbai 400 094, India. \\
$^2$ Human Resources Development Section, Raja Ramanna Centre for Advanced Technology, 
Indore -452013, India.} 

\begin{abstract}
A comprehensive study on the evolution of Stoner factor with doping concentration for various doped 122 systems 
(like BaFe$_2$As$_2$, SrFe$_2$As$_2$) of Fe-based superconductors is presented. Our first principles 
electronic structure calculations reveal that 
for Co/Ru (electron or iso-electronic) doping at Fe sites or P doping at As sites result in a reduction of Stoner factor with 
increasing doping concentration. On the contrary, in case of Na/K (hole) doping at the Ba sites, Stoner factor is 
enhanced for higher doping concentrations. This may be considered as an indicator of elevation of ``magnetic fluctuation''
in these systems. We find that the Stoner factor uniquely follows the variation of 
the pnictide height z$_{As}$/Fe-As bond length with various kinds of doping. 
Our calculated Fermi surfaces explicate the diversities in the behaviour of Stoner 
factors for various doped 122 systems ; larger degree of Fermi surface nesting, larger the value of Stoner factor and vice versa.
\end{abstract}
\maketitle



\vspace{2.5cm}
\section{Introduction}
\label{Introduction}

Invention of superconductivity in Fe-based compounds not only introduced a drastic change in the 
belief that Fe is inimical to superconductivity due to the presence of strong local magnetic moment 
associated with Fe atom but also recreate a glimpse of hope for developing a complete generalized
theory of high temperature superconductivity. One of the most interesting aspects of these Fe-based 
superconductors (SCs) from a fundamental point of view, is that superconductivity may arise from 
magnetic fluctuation or orbital fluctuation \cite{Stewart,Hosono,Fernandes,Wu,Kontani}. However, the glue to the 
electron-electron attraction or more precisely 
pairing mechanism in these Fe-based SCs is far from being settled \cite{Kontani,Mazin,Chubukov1,Jalcom}. Phase diagrams of Fe-based SCs 
establish the manifestation of various exotic phases like spin density wave (SDW), orbital order, 
nematic order, structural transition {\it etc.} \cite{Avci,Sn,Lucarelli,XFWang,Ni,Thaler,Nandi,Kasahara}. 
Various physical properties including superconductivity 
of these Fe-based SCs are very much sensitive to temperature, pressure as well as doping concentration as 
evident from phase diagrams \cite{Mani,Park,Sefat,sust}. Influence of structural parameters on the 
superconducting as well as on other 
exotic phases is well established through extensive theoretical and experimental investigations \cite{zAs,acta,sust}. 
Magnetic (antiferromagnetic spin) fluctuation in high T$_c$ cuprates provide pairing in 
d$_{x^2-y^2}$ channel whereas in Fe-based superconductors it is believed to provide s$^{\pm}$ pairing 
symmetry \cite{Mazin}. The quantity that may be used as an indicator of magnetic fluctuation in metallic systems, is 
Stoner factor. Although Stoner theory is formulated for ferromagnetic instability, it can 
also be applied for anti-ferromagnetic systems \cite{MazinStoner}.\\

Broadly, there exists about six invented families of Fe-based SCs \cite{Stewart}.
Among all these families 122 family is the most studied one both theoretically and experimentally 
largely because of availability of high quality single crystals in this family.
The generalized chemical formula of 122 family is MFe$_2$X$_2$, where M$=$Ba, K, Ca, Sr and X$=$As, P.
The phase diagrams of 122 family 
exhibit a large number of diversities in the physical properties \cite{Stewart}. For example, superconductivity 
emerges in the electron doped systems ({\it e.g.,} replacing
Co by Fe) with a very small doping concentration ($\sim$$2$\%) \cite{Ni,Thaler,Nandi} 
whereas the same arises in hole doped systems with higher doping concentration 
($\sim$$25$\%) \cite{Avci}. On the other hand, for the case of iso-electronic Ru doping at the Fe site, superconductivity 
appears at about 50\% Ru doping concentration \cite{Sharma1,Fan} which is quite high compared to other doped 122 
systems. Another qualitatively distinct diversity in the phase diagram of 122 system is observed in case of Na 
doped Ba122 system \cite{Avci2}. Avci {\it et al.,} displayed the appearance of an additional C$_4$ (tetragonal) phase well inside 
the orthorhombic phase in Ba$_{1-x}$Na$_x$Fe$_2$As$_2$ system at the boundary between superconductivity and SDW 
(stripe antiferromagnetism) which is different from the previous observations of 
re-entrant tetragonal phase in electron-doped compounds \cite{Nandi}
 and has many important consequences. The role of various structural parameters specially 
z$_{As}$ (fractional z co-ordinate of As atom) or anion height (height of the As atom from the Fe plane) 
in superconductivity, magnetism and structural transition is indispensable 
\cite{Mizuguchi,Ganguli,Lee,Zhao,Shirage,acta,sust}. However the inter-relationship 
among superconductivity, magnetic order, nematic order, orbital order and structural transition is 
under thorough investigations both theoretically as well as experimentally \cite{arxiv,Fernandes,Kontani,Luo1, Luo2}. 
The presence of magnetic fluctuation in these Fe-based systems 
is probed by magnetic susceptibility ($\chi$). A large number of theoretical and experimental works 
are available in the literature addressing the role of magnetic fluctuation in superconductivity and other 
observed physical properties of Fe-based SCs \cite{Taylor,arxiv}. An important finding of the present work is that the Stoner factor of various 122 material (that may model magnetic fluctuation) varies with doping in the same fashion as that of the pnictide height or z$_{As}$. 

Density functional theory (DFT), in its most simple implementation [the local-(spin)-density-approximation
(L(S)DA)] is in principle an exact theory to estimate the ground state properties of any system.
It is now well established that LSDA and generalized gradient approximation (GGA) approach, 
which does not take into account the effect of near-critical fluctuations on
the long-range magnetism, fails to predict accurate electronic structures for itinerant magnetic system near 
quantum critical point. A significant amount of effort has been made to introduce these fluctuations within first principles 
DFT method \cite{MazinSpin,Ferber}.
Stoner factor is a parameter 
which mimic the presence of magnetic instability in a system and can be 
calculated within LSDA approach as shown in ref \cite{MazinStoner}. 
However no thorough comprehensive investigations, either theoretically or experimentally exist as to how 
Stoner factor is being modified due to various kinds of dopings and  temperature.
In this paper we present a comprehensive picture of Stoner factor for various 122 Fe-based systems as a function of doping concentration 
from first principles simulations.\\

\begin{figure}
 \centering
 \includegraphics[height=6.0cm,width=7.0cm]{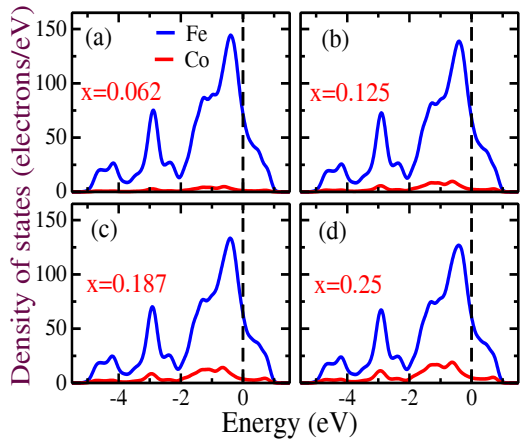}
 \caption{Calculated atom projected density of states within super-cell method 
  for Fe and Co atoms in Co doped (electron doping) Ba122 system (BaFe$_{2-x}$Co$_x$As$_2$) 
  in the tetragonal phase for various doping concentrations, indicated in the figure.}
\label{sc-Co}
\end{figure}
 We calculate Stoner factor from first principles study of density of states (DOS) for these 122 
 compounds as a function of doping concentrations. We have found that Stoner factors 
 get modified qualitatively differently for different kinds of doping. 
 For example, in case of electron doping (Co doping at the Fe site) and iso-valent doping
 (Ru doping at Fe site as well as P doping at As site), Stoner factor gradually decreases with increasing 
 doping concentration. On the other hand, exactly opposite trend in the behaviour of Stoner factor 
 is observed in case of hole doped systems (K/Na doping at the Ba site).  
 This diversity in the behaviour of Stoner factor is then analysed in the light of 
 calculated nature of the FSs of these systems for various doped 
 cases. The method of calculations, including the evaluation of Stoner factor for various doped systems 
 require magnitude of DOS at the Fermi level which are obtained through detailed first principles simulation and 
 are discussed in the forthcoming theoretical method section. 
 An important speciality of our simulation results is that we use exact experimentally determined structural parameters 
 of various doped systems as input which are being kept intact. As a consequence of that, presented results 
 are very much realistic.
 In the succeeding section we present the variation of 
 Stoner factor with doping concentration for many doped materials of 122 family of Fe-based SCs. 
 Modifications of the FSs are also depicted to elucidate the observed anomalies in the 
 behaviour of Stoner factor with doping. In the last section, we finally conclude and summarize our results. 
\begin{figure}
 \centering
 \includegraphics[height=8.0cm,width=7.5cm]{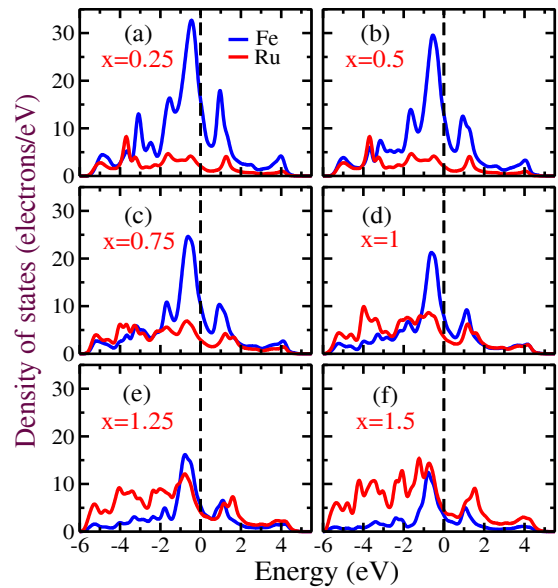}
 \caption{Calculated atom projected density of states within super-cell method 
   for Fe and Ru atoms in Ru doped (iso-electronic doping) Sr122 system (SrFe$_{2-x}$Ru$_x$As$_2$) 
   in the tetragonal phase for various doping concentrations, indicated in the figure.}
\label{sc-Sr}
\end{figure}

\section{Theoretical Methods}
\label{Method}
Our first principles electronic structure calculations are performed by implementing ultrasoft 
pseudopotential with plane wave basis set based on density functional theory \cite{CASTEP}. Electronic 
exchange correlation is treated under the generalised gradient approximation (GGA) using 
Perdew-Burke-Ernzerhof (PBE) functional \cite{PBE}. It should be mentioned here that density funtional 
theory within local density approximation (LDA) as well as generalized gradient approximation (GGA) was
unable to optimize the experimental value of $z_{As}$ (fractional co-ordinate of As)
with desired accuracy \cite{acta,sust,DJSingh,Zhang,Yin}. In fact, the optimized value of $z_{As}$ 
is about 0.1 to 0.15 {\AA} smaller than that of the experimental one. The origin of this discrepancy 
is the presence of strong magnetic fluctuation, associated with Fe atoms
in these materials \cite{Mazin2}. Electronic structures (band structure, Fermi surface {\it etc.}),
calculated using optimized lattice parameters ($a$, $b$, $c$ and $z_{As}$) do not resemble with that of the 
experimentally measured one \cite{pla,zAs}. This insist us to employ experimental 
lattice parameters {\it i.e.,} $a$, $b$, $c$ and $z_{As}$ 
\cite{Avci,thesis,acta,Sefat,Ni,Avci3,Allred} instead of geometry optimized 
(total energy minimization) lattice parameters as inputs of our first principles 
electronic structure calculations. We use experimental 
orthorhombic (low temperature) as well as tetragonal (high temperature) lattice parameters 
$a$, $b$, $c$ and $z_{As}$ as input of our {\it ab-initio} electronic structure 
calculations \cite{Avci,thesis,acta,Sefat,Ni,Avci3,Allred}. Various modern X-ray diffraction techniques e.g., using Synchrotrons radiation sources etc. that determines crystallographic information at different external perturbations (like temperature, chemical doping etc.) are essentially result of diffraction from various atomic charge densities (Bragg's diffraction). Considering experimentally determined structural parameters at different temperatures, doping as input thus in turn provides temperature/doping dependent correct densities to our first principles calculation. These input structural parameters are kept fixed through out the calculation for a fixed doping. This is how we use the DFT formalism to bring out realistic doping dependent observable with the help of experimental input (energy being functional of electron density
 E $\equiv E[\rho(r,T,x)]\equiv E[\rho (a(x,T), b(x,T),c(x,T))$]. The main effect on the electronic structure from finite temperature/doping is the underlying crystal structure, and the average crystal structure can usually be reliably determined from the diffraction experiment at a given doping.

In order to dope the system, we use two theoretical methods: (1) Virtual crystal approximation (VCA) \cite{Bellaiche} 
(2) Super-cell. These two methods are used to prepare various doped Ba122 and Sr122 systems as per requirement. 
For example, we use super-cell method for calculating the DOS as well as the FSs of Ru doped 122 systems because 
the VCA method is unable to produce accurate electronic structures in the 
higher doping regime for these systems. On the contrary, we employed 
VCA method to dope K, Na and P in Ba122 system. 
The detailed discussion about the application of VCA and super-cell methods 
for various doped 122 systems can be found in ref \cite{vca}. In this work, we use the VCA method, 
developed by Bellaiche and Vanderbilt \cite{Bellaiche} based on weighted averaging of pseudopotentials. 
It should be noted that although 
DOS of Co doped 122 systems are calculated using super-cell method, FSs are 
calculated within VCA method to handle the small percentage of doping in the primitive unit cell. Moreover, size of 
the super-cells are different for different doping cases. For example, in Ru doped Ba122 and Sr122 systems, we use 
$\sqrt{2}\times\sqrt{2}\times 1$ super-cells, whereas $2\times 2 \times 1$ super-cell is used 
for Co doped 122 materials.
Non-spin-polarized and spin polarised single point energy calculations 
were carried out for tetragonal phase with space group symmetry I4/mmm (No. 139) and 
orthorhombic phase with space group symmetry Fmmm (No. 69) respectively 
using ultrasoft pseudo-potentials and plane-wave basis 
set with energy cut off 500 eV and higher as well as
self-consistent field (SCF) tolerance as $10^{-6}$ eV/atom. 
Brillouin zone is sampled in the k space within Monkhorst–Pack scheme
and grid size for SCF calculation is chosen as per requirement of the calculation 
for various systems and approaches. For simulating Fermi surfaces, grid size of SCF calculation is chosen as
$26\times 26\times 31$. LDA+U calculations have also been performed 
 for orthorhombic Na and P doped Ba122 systems with SDW spin configuration \cite{SDW}, where 
 doping has been implemented within super-cell method.
Stoner factor of a compound like doped 122, can be defined as
$I^{Fe}\times[N^{Fe}(E_F)]^2+I^{Ru/Co}\times[N^{Ru/Co}(E_F)]^2$, 
where $N^{Fe}(E_F)$ and $N^{Ru/Co}(E_F)$ are the density of states
at the Fermi level from Fe and Ru/Co atoms respectively \cite{Stoner,MazinStoner}. 
The values of Stoner parameters $I^{Fe}$ and $I^{Ru/Co}$ are taken 
from ref \cite{Stoner,Yan,Sefat2}.
\begin{figure}
 \centering
 \includegraphics[height=8.0cm,width=7.5cm]{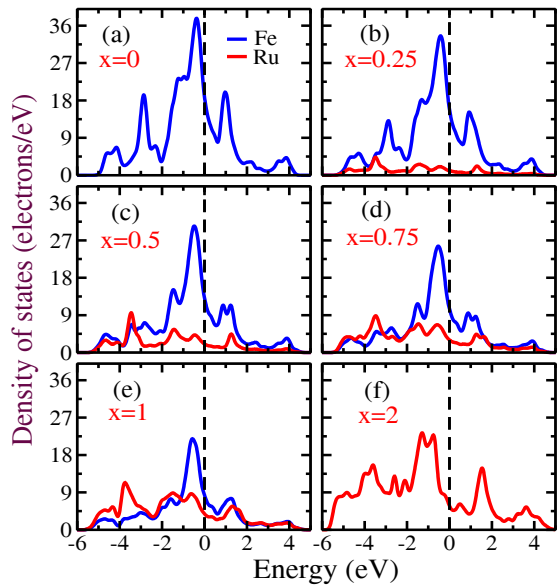}
 \caption{Calculated atom projected density of states within super-cell method 
    for Fe and Ru atoms in Ru doped (iso-electronic doping) Ba122 system (BaFe$_{2-x}$Ru$_x$As$_2$) 
    in the tetragonal phase for various doping concentrations, indicated in the figure.}
\label{sc-Ba}
\end{figure}
\begin{figure}
 \centering
 \includegraphics[height=8.0cm,width=7.5cm]{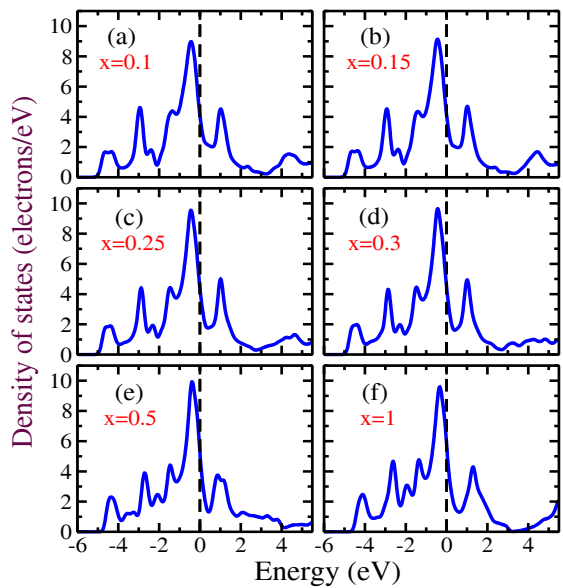}
 \caption{Calculated atom projected density of states within VCA method 
   for Fe atoms in K doped (hole doping) Ba122 system (Ba$_{1-x}$K$_x$Fe$_2$As$_2$) in the 
   orthorhombic phase for various doping concentrations, indicated in the figure.}
\label{vca-K-ortho}
\end{figure}

\section{Results and discussions}

We have studied many doped samples (hole doping, electron doping and iso-valent doping) 
of 122 Fe-based SCs to study the evolution of Stoner 
factor with doping concentration. 
As a representative of hole doped and electron doped systems, we consider Na/K doping (at the Ba site) 
 and Co doping (at the Fe site) of BaFe$_2$As$_2$ (Ba122) systems respectively. Isovalent or iso-electronic doping at 
 Fe site (Ru doping) for BaFe$_2$As$_2$ and SrFe$_2$As$_2$ (Sr122) systems as well as in As site (P doping) 
 for BaFe$_2$As$_2$ system are also considered in our investigation.
In the theoretical method section we have already described the method of calculation 
of Stoner factor of a compound system within density functional theory. 
Virtual crystal approximation (VCA) and super-cell methods are used to mimic various doped 122 
Fe-based superconducting systems. 
We employed experimental lattice parameters $a$, $b$, $c$ and $z_{As}$ in both the
orthorhombic and tetragonal phases for our first 
principles electronic structure (DOS and FSs) calculations. 
It is well known that in 122 Fe-based SCs, DOS at the Fermi level is 
dominated by Fe-d orbital and to some extent by As-p orbitals. It is for this reason, 
the particular case of doping at the Fe site is treated within super-cell method 
in order to calculate the atom projected DOS for Fe and Co/Ru atoms separately 
with desirable precision. It should be mentioned here that 
within VCA method, the atom projected DOS of Fe and Ru/Co atoms can not be estimated 
separately. And for other doping cases (K/Na doping at the Ba site and P doping at As site) 
we employ VCA for implementation of doping (in which Fe atom projected DOS can be obtained). 
We have also employed super-cell method for implementing K/Na/P doping in 122 system, but the 
results of which are not presented here, are very similar to that of the VCA results.
\begin{figure}
 \centering
 \includegraphics[height=8.0cm,width=7.5cm]{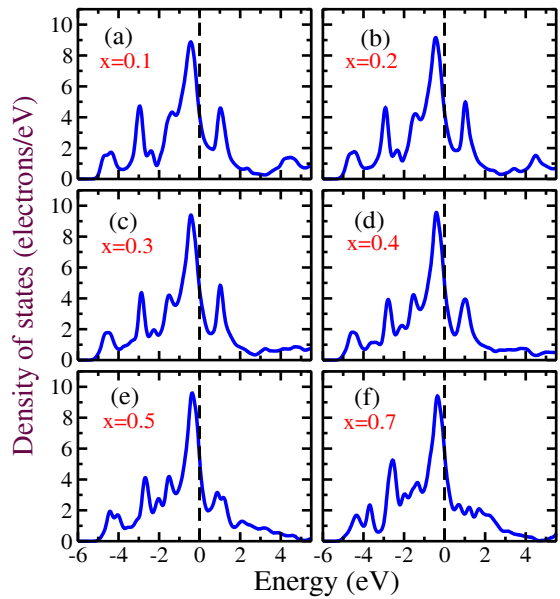}
 \caption{Calculated atom projected density of states within VCA method 
   for Fe atoms in Na doped (hole doping) Ba122 system (Ba$_{1-x}$Na$_x$Fe$_2$As$_2$) in the 
   orthorhombic phase for various doping concentrations, indicated in the figure.}
\label{vca-Na-ortho}
\end{figure}
\begin{figure}
 \centering
 \includegraphics[height=8.0cm,width=7.5cm]{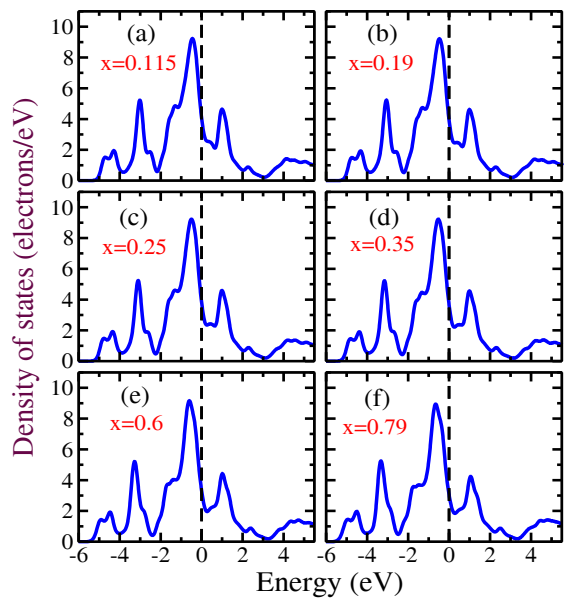}
 \caption{Calculated atom projected density of states within VCA method 
   for Fe atoms in P doped (iso-electronic doping) Ba122 system BaFe$_2$(As$_{1-x}$P$_x$)$_2$ in the 
   orthorhombic phase for various doping concentrations, indicated in the figure.}
\label{vca-P-ortho}
\end{figure}
First, we calculate the atom projected DOS for Co doped Ba122 systems (electron doping) for several 
Co doping concentrations in the tetragonal phase.
In FIG.\ref{sc-Co}, DOS contribution of Fe and Co atoms are separately presented 
for various doping concentrations, indicated in the figures. As discussed earlier, 
for electron doped system (like Co doping at Fe site) superconducting as well as other 
exotic phases appear in the system within 10\% to 15\% doping concentration which 
is smaller compared to that for other doped 122 systems (for example, superconductivity arises in Ru doped Ba122 system at
a very high doping concentration). That is why we restrict our calculation in the low doping regime 
in the case of Co doped Ba122 system. We depict atom projected DOS in the tetragonal phase for Ru doped Sr122 and 
Ba122 systems in FIG.\ref{sc-Sr} and FIG.\ref{sc-Ba} respectively. In those figures, DOS 
contribution of Ru and Fe atoms are presented separately as a function of Ru doping concentration. 
In this case our study is no longer restricted to low doping regime but 
extended our investigation for whole regime of doping concentrations (low to high). 
On the other hand, we exhibit atom projected DOS in the orthorhombic phase for Fe atom in case of 
K, Na and P doped Ba122 systems in FIG.\ref{vca-K-ortho}, FIG.\ref{vca-Na-ortho} and FIG.\ref{vca-P-ortho} 
respectively. Moreover, atom projected DOS in the tetragonal phase for K, Na and P doped Ba122 systems 
are also calculated but not presented here. 
In these cases we employ VCA method for handling doped systems as Fe site is not directly 
affected by K and Na doping at the Ba site (hole doping) and P doping at the As site (iso-electronic doping). 
However, we also calculate Fe atom projected DOS for these systems using super-cell approach. 
We find DOS calculated within super-cell method is qualitatively similar to that 
using VCA method.

\begin{figure}
 \centering
 \includegraphics[height=4.5cm,width=8cm]{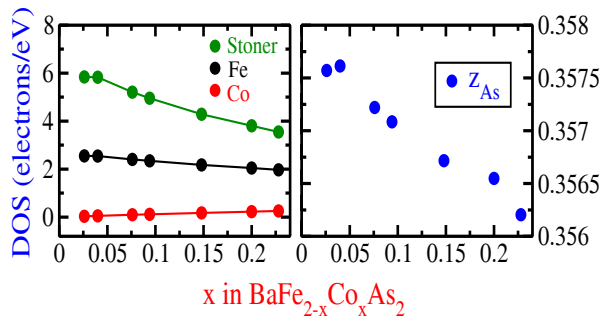}
 \caption{(\textbf{a}) Contrasting nature of Stoner factor (green) of BaFe$_{2-x}$Co$_x$As$_2$ system in the tetragonal phase as a function
 of Co doping concentration. Variation of density of states of Fe (black) and Co (red) as a function of Co doping concentration 
 is also shown. (\textbf{b}) Variation of experimental (also one of the input parameter of our calculation) z$_{As}$
 (fractional z co-ordinate of As atom) with Co doping concentration.}
\label{Stoner-Co}
\end{figure}
\begin{figure}
 \centering
 \includegraphics[height=6.5cm,width=8cm]{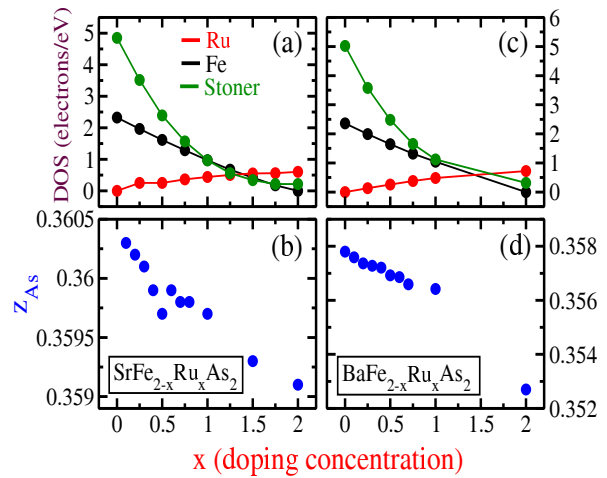}
 \caption{(\textbf{a, c}) Calculated variation of Stoner factor (green) of SrFe$_{2-x}$Ru$_x$As$_2$ and
 BaFe$_{2-x}$Ru$_x$As$_2$ systems in the tetragonal phase as a function of Ru doping concentration 
 respectively. Variation of density of states of Fe (black) and Ru (red) as a function of Ru doping concentration 
 is also shown. (\textbf{b, d)} Variation of experimental (also one of the input parameter of our calculation) z$_{As}$
 (fractional z co-ordinate of As atom) with Ru doping concentration.}
\label{Stoner-Ru}
\end{figure}
\begin{figure}[ht]
 \centering
 \includegraphics[height=6.5cm,width=8cm]{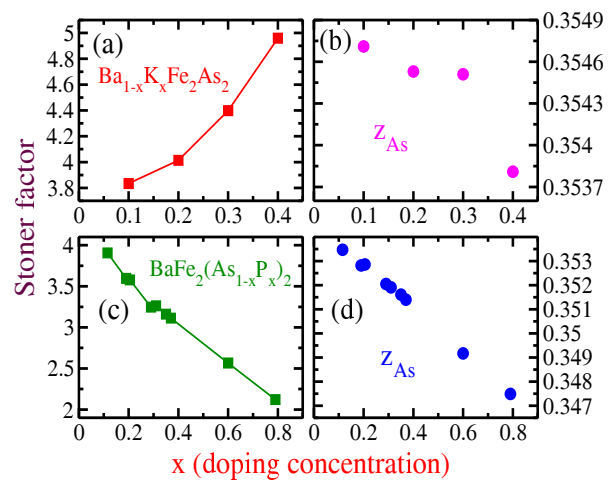}
 \caption{(\textbf{a, c}) Calculated variation of Stoner factor of Ba$_{1-x}$K$_x$Fe$_2$As$_2$ and
   BaFe$_2$(As$_{1-x}$P$_x$)$_2$ systems in the tetragonal phase as a function of K and P doping concentration 
   respectively. (\textbf{b, d}) Variation of experimental z$_{As}$ (also one of the input parameter of our calculation)
   (fractional z co-ordinate of As atom) with doping concentration.}
\label{Stoner-tetra}
\end{figure}
\begin{figure}
 \centering
 \includegraphics[height=7.0cm,width=7.5cm]{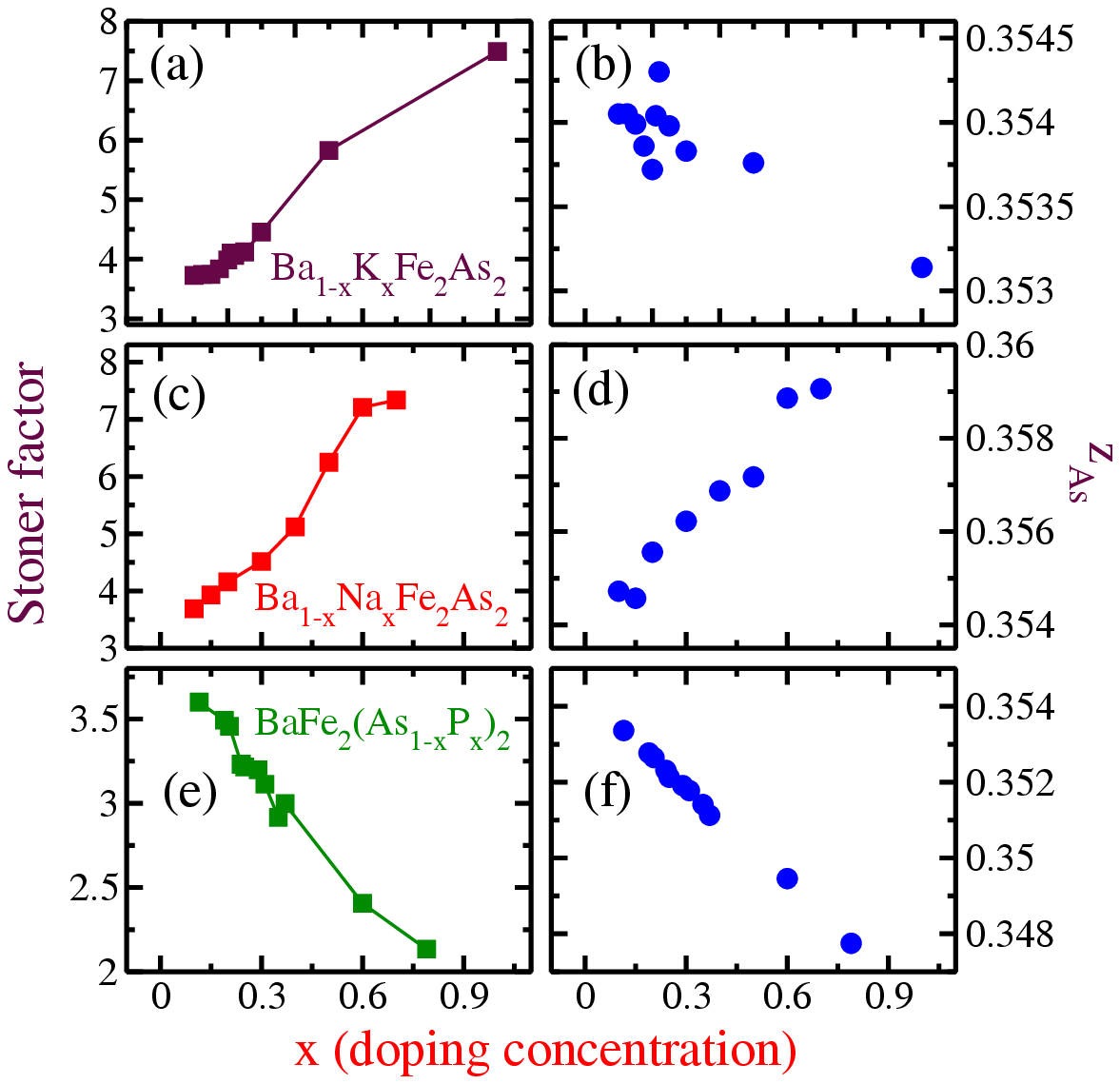}
 \caption{(\textbf{a, c, e}) Calculated variation of Stoner factor of Ba$_{1-x}$K$_x$Fe$_2$As$_2$, Ba$_{1-x}$Na$_x$Fe$_2$As$_2$ and
  BaFe$_2$(As$_{1-x}$P$_x$)$_2$ systems in the orthorhombic phase as a function of K, Na and P doping concentration 
  respectively. (\textbf{b, d, f}) Variation of experimental (also one of the input parameter of our calculation) z$_{As}$
  (fractional z co-ordinate of As atom) with doping concentration.}
\label{Stoner-P-k-Na}
\end{figure}
\begin{figure}[ht]
 \centering
 \includegraphics[height=4.5cm,width=8cm]{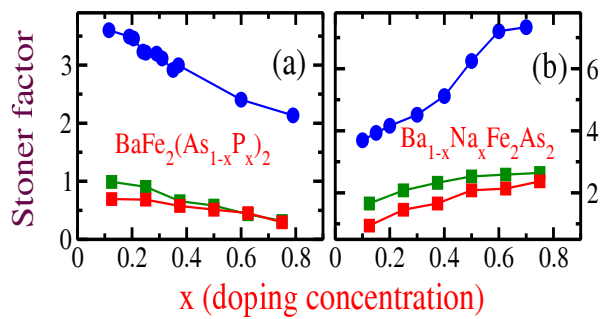}
 \caption{Calculated variation of Stoner factor of (\textbf{a}) BaFe$_2$(As$_{1-x}$P$_x$)$_2$ and (\textbf{b}) Ba$_{1-x}$Na$_x$Fe$_2$As$_2$ 
   systems in the orthorhombic phase as a function doping concentration for different U values. 
   Blue circles are calculated Stoner factor with U=0. Green and red square are the 
   calculated Stoner factor for U=0.5 and U=1 respectively.}
\label{Stoner-U}
\end{figure}
\begin{figure}
 \centering
 \includegraphics[height=7.0cm,width=7.5cm]{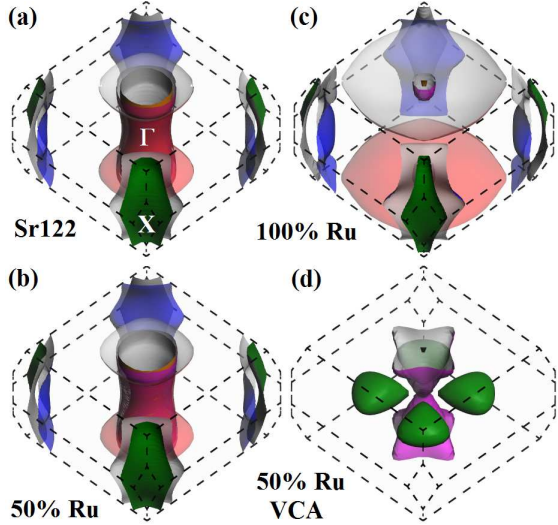}
 \caption{Calculated FSs of (\textbf{a}) undoped Sr122 (\textbf{b}) 50\% and (\textbf{c}) 100\% Ru doped Sr122 system using super-cell method. 
   (\textbf{d}) Calculated FSs of 50\% Ru doped Sr122 system within VCA method.
   Doping concentration is also indicated in the figure.
   Different colours are used to indicate different FSs. There are three hole like FSs around $\Gamma$ point (centre of the BZ) 
   and two electron like FSs around X point (four corners of the BZ). With 50\% Ru doping there is no 
   significant change in the FS topology but most of the FSs of 100\% Ru doped Sr122 system are completely 3D like.
   Calculated FSs for 50\% Ru doped Sr122 system within VCA method, do not matches with experimental observations.}
\label{FS-Ru}
\end{figure}

Using this atom projected DOS, we calculate Stoner factor using the 
method, described in the theoretical method section for various doped 122 systems.
In FIG.\ref{Stoner-Co}a, we display the Stoner factor as well as atom  projected DOS of Fe and 
Co atoms at Fermi level as a function of Co doping concentration in the tetragonal phase. 
In FIG.\ref{Stoner-Co}b we provide estimated z$_{As}$ (also use as an input of our 
first principles calculation) as a function of Co doping concentration.
We observe that with increasing Co doping concentration, both the Stoner factor as well as z$_{As}$ 
decreases. We have also calculated the FSs of Co doped Ba122 system within VCA method. 
Our calculated FSs also provide a clear indication of more 3D like FSs with higher Co doping 
concentration. This 3D like FSs work against nesting and results in suppression of magnetic 
order (SDW). Since Stoner factor is intimately related to magnetic fluctuation or magnetic instability, reduction of 
Stoner factor with increasing Co doping concentration gives a clear indication of depletion of possible  
``magnetic fluctuation'' in the system with the introduction of Co. This is consistent with the evolution 
of calculated FSs as a function of doping. Now we move to the case of iso-electronic doping at Fe site 
(Ru doping at Fe site of Ba122 and Sr122 systems). For both the systems with increasing doping concentration 
z$_{As}$ decreases as shown in FIG.\ref{Stoner-Ru}b and FIG.\ref{Stoner-Ru}d. In FIG.\ref{Stoner-Ru}a and FIG.\ref{Stoner-Ru}c 
atom projected DOS at the Fermi level for Ru and Fe atoms as well as Stoner 
factor are shown as a function of Ru doping concentration for tetragonal Sr122 and Ba122 systems respectively. 
For both the cases Stoner factor decreases with increasing Ru doping concentration in a very similar manner. 
We have also performed the FS calculation for Ru doped Sr122 system within supercell method for 
50\% Ru doping case. We find that for 122 system 50\% Ru doping modifies the FSs from 2D like 
 to more 3D like, which is consistent with previous theoretical and experimental investigations \cite{pla,Xu}. 
 This in-turn degrades the nesting of FSs and triggers superconductivity in the systems by means of suppression 
 of SDW order or magnetic order. This again creates an impression of curtailment of magnetic order 
 with increasing Ru doping concentration, which is also consistent with the calculated variation of Stoner factor 
 with Ru doping. As far as our FS calculation is concerned, upto 50\% Ru doping 
 in Sr122 system, no significant moderation is observed in the FS topology in contrast to Ba122 system. 
 However SrRu$_2$As$_2$ has completely 3D like FSs just like BaRu$_2$As$_2$ system, presented in FIG.\ref{FS-Ru}c. 
 In FIG.\ref{FS-Ru}d we depict the calculated FSs of 50\% Ru doped Sr122 system within VCA method. 
 The topology of FSs calculated within VCA method, certainly do not match with 
 the experimental one as reported in literature \cite{pla}.\\
\begin{figure}
 \centering
 \includegraphics[height=7.0cm,width=7.5cm]{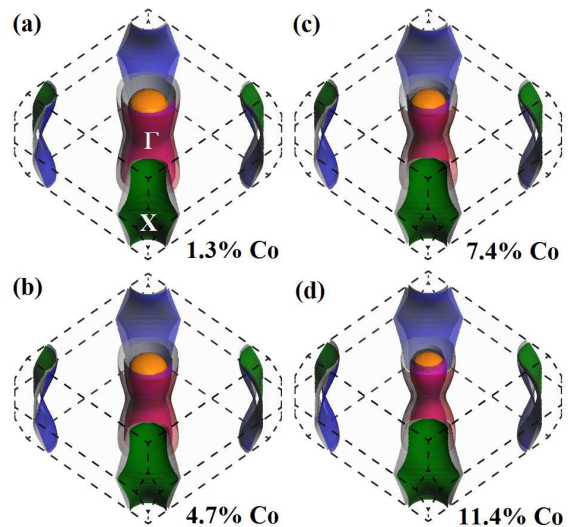}
 \caption{Calculated FSs of various Co doped Ba122 systems. Doping concentration is also indicated in the figure.
  Different colours are used to indicate different FSs. There are three hole like FSs around $\Gamma$ point (centre of the BZ) 
  and two electron like FSs around X point (four corners of the BZ). With increasing doping concentration
  hole like FSs shrink and electron like FSs expand (electron doping). Also for higher doping concentrations 
  the FSs are more 3D like.}
\label{FS-Co}
\end{figure}
\begin{figure}
 \centering
 \includegraphics[height=7.0cm,width=7.5cm]{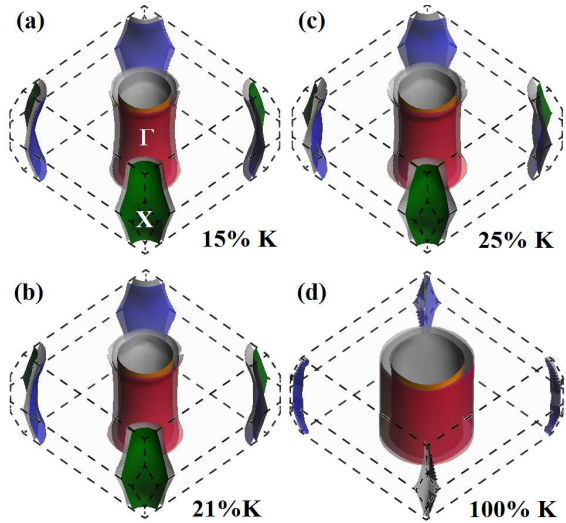}
 \caption{Calculated FSs of various K doped Ba122 systems. Doping concentration is also indicated in the figure.
    Different colours are used to indicate different FSs. There are three hole like FSs around $\Gamma$ point (centre of the BZ) 
    and two electron like FSs around X point (four corners of the BZ). With increasing doping concentration
    hole like FSs expand and electron like FSs shrink (hole doping). Also for higher doping concentrations 
    the FSs are more 2D like. For 100\% K doped system electron like FSs almost disappear.}
\label{FS-K}
\end{figure}

\begin{figure}
 \centering
 \includegraphics[height=7.0cm,width=7.5cm]{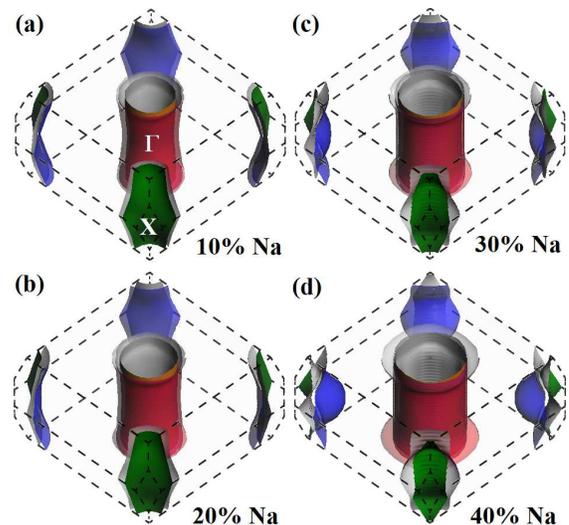}
 \caption{Calculated FSs of various Na doped Ba122 systems. Doping concentration is also indicated in the figure.
   Different colours are used to indicate different FSs. There are three hole like FSs around $\Gamma$ point (centre of the BZ) 
   and two electron like FSs around X point (four corners of the BZ). With increasing doping concentration
   hole like FSs expand and electron like FSs shrink (hole doping). Also for higher doping concentrations 
   the FSs are more 2D like.}
\label{FS-Na}
\end{figure}

\begin{figure}[ht]
 \centering
 \includegraphics[height=7.0cm,width=7.5cm]{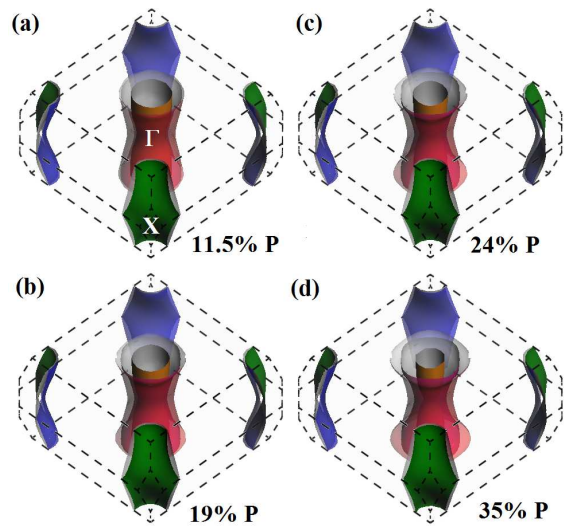}
 \caption{Calculated FSs of various P doped Ba122 systems. Doping concentration is also indicated in the figure.
   Different colours are used to indicate different FSs. There are three hole like FSs around $\Gamma$ point (centre of the BZ) 
   and two electron like FSs around X point (four corners of the BZ). With increasing doping concentration
   the FSs are more 3D like.}
\label{FS-P}
\end{figure}

Next we calculate the evolution of Stoner factor for K, Na and P doped Ba122 system with increasing doping concentration. 
In FIG.\ref{Stoner-tetra}a, c the variation of Stoner factor as a function of doping concentration 
has been shown for K and P doped tetragonal Ba122 systems respectively. In FIG.\ref{Stoner-tetra}b, \ref{Stoner-tetra}f the variation 
of z$_{As}$ with doping concentration is also presented for the above mentioned systems respectively. 
Calculated evaluation of Stoner factors in the orthorhombic Ba122 systems as a function of K, Na and P doping concentrations 
also show similar trends as that of the tetragonal K, Na and P doped Ba122 systems, shown in FIG.\ref{Stoner-P-k-Na}.
In FIG.\ref{Stoner-P-k-Na}b and \ref{Stoner-P-k-Na}f, the variation 
of z$_{As}$ with doping concentration is also presented for the same systems respectively. 
In case of K and Na doped systems (hole doped) Stoner factor increases with the 
increase of doping concentration which also corroborates well with 
the nature of the calculated FSs, shown in FIG.\ref{FS-K} and \ref{FS-Na}. 
Although exactly opposite trends in the behaviour of z$_{As}$ with doping concentration (experimental) is observed 
in case of K and Na doping (see FIG\ref{Stoner-P-k-Na}b and d) which makes the 
Na doped system different from other hole doped 122 systems. For P doped system both z$_{As}$ and 
Stoner factor decreases with increasing doping concentration, which indicate reduction of magnetic instability with P doping.
So in general, the Stoner factors of various 122 systems follow the variation of z$_{As}$ with doping except in the 
case of K doped Ba122 system where the variation of z$_{As}$ with doping is not well defined. 
Experimental data of z$_{As}$ for K doped Ba122 is not as consistent (probably accurate) 
compared to that for the other doped 122 family of compounds (see FIG. \ref{Stoner-P-k-Na}b, d).
Since our theoretical calculation is based on the experimental crystallographic data as input, 
the calculated electronic structure differs from general observation (actually the experimental 
inconsistency in z$_{As}$ with doping is reflected in the calculated results).
In case of Ba$_{1-x}$K$_x$Fe$_2$As$_2$, thus, comparison of Stoner factor with z$_{As}$ would not be 
a correct quantity to look into but the Fe-As bond length as a function of doping concentration. 
Actually, Stoner factor universally follows the variation of Fe-As bond length for various types of doping. 
In case of hole doping, Stoner factor inversely follows Fe-As bond length with the variation of 
doping concentration. On the other hand, in the case of electron doping, 
Stoner factor follows the doping dependent behaviour of Fe-As bond length. 
In order to understand universality in the behaviour of Stoner factor and z$_{As}$/Fe-As bond length of hole doped Ba122,
 we shed more light into the other structural parameters. To elucidate 
 the inconsistency in the behaviour of Stoner factor and z$_{As}$ in K doped Ba122, 
 we introduce in plane and out of plane As-As distances which also play 
 very important and crucial role. Size of the Ba atom (2.22\AA) [atomic radius in metallic bonding] is larger 
 than Na atom (1.86\AA) but smaller than K atom (2.27\AA). Because of this reason, with the substitution of K atom in place of Ba atom, out of plane As-As distance increases and exactly opposite behaviour in the out 
 of plane As-As distance is observed in the case of Na substitution at Ba site. 
 It should also be noted that the c-axis increases with the increasing doping 
 concentration for both Na and K doping (in case of Na doped Ba122 materials 
 c-axis increases with doping up to certain doping concentration and after that it decreases). 
 This qualitatively explains the observed behaviour of z$_{As}$ with doping 
 for K and Na doped Ba122 systems \cite{Sen2017}. 
 But in both the cases Fe-As bond length  decreases with increasing doping concentration.
Calculated Stoner factors for K/Na/P doped 122 systems using super-cell approach also follow the same 
trends as that of the calculated one within VCA method. 
Calculated FSs also become more and more 3D like with increasing P doping concentration, which is consistent 
with the suppression of magnetic order as observed in experiments (see FIG.\ref{FS-P}). 
On the contrary to all other cases of doping, hole doping makes the FSs more 2D like which actually helps 
having larger degree of nesting of FSs. This increasing trend of Stoner factor with doping, may be related to 
the more 2D like FSs of the hole doped 122 systems that enhances nesting segments. 
In all these 122 systems, superconductivity emerges into the system with certain 
percentage of doping concentration (small or large). On the contrary to 
other 122 Fe-based SCs, Stoner factor of hole doped 122 systems (K/Na doped) display distinctly different 
behaviour. In these cases only, the Stoner factor increases with increasing doping concentration 
which indicate that, these systems become more and more unstable against magnetic instability. 
So in these hole doped 122 systems, superconductivity arises at the same moment when magnetic 
instability also dominates. 
We have also studied the behaviour of Stoner factor in presence of electron correlation which can 
be introduced through LDA+U calculation within DFT. Since Fe based SCs specially 122 systems 
are weakly correlated, we restrict our LDA+U calculation for a small 
value of U, which defines the strength of correlation (U=0.5 and U=1). 
In FIG.\ref{Stoner-U} the variation of Stoner factor with doping concentration for various U values 
(U=0.5 and U=1) are depicted for  BaFe$_2$(As$_{1-x}$P$_x$)$_2$ and Ba$_{1-x}$Na$_x$Fe$_2$As$_2$ systems 
in the orthorhombic phase. It is to be noted here that with the introduction of U, Stoner factor decreases for both the 
cases but shows very similar trends with doping concentration as that in the case of U=0. 
This study (of LDA+U) has also been extended to all other materials presented in this work.
The reduction of Stoner factor with correlation is consistent with the current literature \cite{MazinStoner}.
It should also be noted here that K-doped 
Ba122 systems have the highest superconducting transition temperature (T$_c$) among all other 
Fe-based SCs in ambient condition. This may give some impression that 
superconductivity in these systems is intimately related to magnetic order and magnetic fluctuations. 
On the whole, our study provides a clear view about the variation of Stoner factor as a function 
of doping for all variety of doped 122 Fe-based SCs. We also explain the observed diversities in the behaviour 
of various doped 122 systems by presenting calculated FSs.


The phase diagram of the iron pnictides reveals that the superconducting transition temperature T$_c$ usually shows a dome-like relation with doping. Should the Stoner factor vary with doping in the fashion as Tc in phase diagrams ? One should be aware of the fact that both temperature and doping plays a very important role in any phase diagram. In fact in phase diagram, both temperature and doping dependencies of various phases like spin density wave, superconducting phase \textit{etc.}, are known for various Fe based superconductors. However, in our first principles electronic structure calculations, we use experimentally measured doping dependent lattice parameters ($a$, $b$, $c$, z$_{As}$) at a fixed temperature as input. That is why our calculated Stoner factor does not contain any temperature dependencies (only doping dependency). In general, z$_{As}$ (as well as other structural parameters) has a very sensitive temperature dependencies. For example, in our earlier work we have shown in Ru doped Ba122 system, structural parameters including z$_{As}$ are highly sensitive to temperature \cite{acta}. 
When such a temperature dependency was included in the electronic structure calculation by us in the 
past, the temperature dependency of Stoner factor did reproduce the experimental magnetic fluctuation behaviour \cite{arxiv,Matan}.
In the current perspective where we discuss detailed behaviour of Stoner factor over a wide doping ranges for a number of doped-122 family, clean crystallographic data, particularly z$_{As}$ as a function of both, doping and temperature is not available. As a result such a calculation of Stoner factor as a function doping as well as temperature is only a future possibility.
That is why the Stoner factor does not show dome like behaviour.

\section{Conclusion}

It is well documented that the magnetic fluctuation plays a significant role in superconductivity 
of Fe-based SCs. Stoner factor is the precursor of magnetic fluctuation or magnetic instability in a system. 
We have presented in detail the modification of Stoner factor with doping concentration for 
various doped 122 Fe-based SCs. VCA as well as super-cell methods are employed to dope these materials 
theoretically to calculate DOS and FSs of doped 122 compounds. Diversities in the behaviour of 
Stoner factor with varying doping concentrations for various nature of doped 122 systems are discussed. 
In case of Ru (iso-electronic)/Co (electron) doped systems, 
Stoner factor decreases with increasing Ru/Co concentration which is compatible with the calculated FSs. 
Same trends in the behaviour of Stoner factor and FSs are also observed in the case of other iso-electronic doping,
P doping at the As site. On the other hand, in case of hole doped systems (Na/K doping at the  Ba site) 
Stoner factor increases with increasing doping concentration which is also consistent with the 
corresponding modifications in the calculated FSs for various doping concentrations. Remarkably, we find that the Stoner factor follows the 
variation as that of the z$_{As}$ with doping. Larger the pnictide height larger is the Stoner factor as well as larger degree of the Fermi surface nesting, larger the value of Stoner factor  and vice versa.
 Sensitive dependence of superconducting T$_c$ with anion 
height was established earlier \cite{Mizuguchi} but our present work clearly establishes the same for Stoner factor or magnetic fluctuation. 
As a whole this work provide a comprehensive study of Stoner 
factor (magnetic instability) for various doped 122 Fe-based SCs.\\
\\ 

{\bf Acknowledgements} \\

We thank Dr. A. Bharathi and Dr. A. K. Sinha for discussion on experimental 
aspects. We thank Dr. P. A. Naik and Dr. P. D. Gupta for their encouragement in this work. One of 
us (SS) acknowledges the HBNI, RRCAT for financial support and encouragements.\\ 

{\bf Additional information} \\

{Competing financial interests:} The authors declare no competing financial interests. \\





\end{document}